\documentclass[11pt,a4paper]{article}       
\usepackage[affil-it]{authblk}

\usepackage{braket,amsfonts, amsthm}

\usepackage{array}

\usepackage{pgfplots}

\usepackage{algorithm}
\usepackage{algorithmic}
\usepackage{cite}
\usepackage{dsfont}
\usepackage{graphicx,epstopdf}

\newtheorem{thrm}{Theorem}[section]

\newtheorem{proposition}[thrm]{Proposition}

\newtheorem{definition}[thrm]{Definition}

\newtheorem{xmpl}[thrm]{Example}

\newcommand{\Q}{\mathcal Q}
\renewcommand{\S}{\mathcal S}
\usepackage{graphics} 
\usepackage{dsfont}
\usepackage{epsfig} 
\usepackage{mathptmx} 
\usepackage{times} 
\usepackage{amsmath} 
\usepackage{amssymb}  
\usepackage{mathtools}
\usepackage{siunitx}
\usepackage{subcaption}
\usepackage{tikz}
\usetikzlibrary{arrows, automata, backgrounds, calc, decorations.pathmorphing, decorations.shapes, fit, intersections, matrix, petri, plotmarks, positioning, shapes, snakes, through, trees}
\usepackage{xspace}
\usepackage{bold-extra}
\usepackage{bbm}
\usepackage[most]{tcolorbox}

\usepackage[skip=1pt]{caption}

\colorlet{texcscolor}{blue!50!black}
\colorlet{texemcolor}{red!70!black}
\colorlet{texpreamble}{red!70!black}
\colorlet{codebackground}{black!25!white!25}

\newcommand{\Real}{\mathbb{R}}

\newcommand{\A}{\mathcal A}
\newcommand{\B}{\mathcal B}
\newcommand{\D}{\mathcal D}
\newcommand{\V}{\mathcal V}

\tikzset{%
    state/.style={circle,fill=none,draw=black,text=black, minimum size = 29pt},
    const/.style={circle,fill=red!40,draw=black,text=black, minimum size = 29pt},
    triang/.style={fill=none,draw,inner sep=0pt,label distance=2pt,minimum size=2pt,circle}
}

\title{\LARGE \bf
A Dynamic Programming Approach for Road Traffic Estimation
}

\author{Mattia Laurini, Irene Saccani, Stefano Ardizzoni, Luca Consolini, and Marco Locatelli
\thanks{All authors are with the University of Parma, Department of Engineering and Architecture,
Parco Area delle Scienze 181/A, 43124 Parma, Italy. E-mails:
{\tt\scriptsize \{mattia.laurini, irene.saccani, stefano.ardizzoni, luca.consolini, marco.locatelli\}@unipr.it}.}}%
\date{}

\begin{document}

\maketitle

\begin{abstract}
We consider a road network represented by a directed graph.
We assume to collect many measurements of traffic flows on all the network arcs, or on a subset of them.
We assume that the users are divided into different groups.
Each group follows a different path.
The flows of all user groups are modeled as a set of independent Poisson processes.
Our focus is estimating the paths followed by each user group, and the means of the associated Poisson processes. 
We present a possible solution based on a Dynamic Programming algorithm.
The method relies on the knowledge of high order cumulants.
We discuss the theoretical properties of the introduced method.
Finally, we present some numerical tests on well--known benchmark networks, using synthetic data.
\end{abstract}

\section{Introduction}
Many municipalities are equipped with systems for measuring traffic flows at main roads and intersections.
This data collection is allowed by different types of sensors, such as inductive loop detectors, video detection systems, microwave radar sensors, or LIDAR sensors.
We address the problem of reconstructing the paths traveled by users based on available flow measurements.

More precisely, we represent a road network as a directed graph and assume that several traffic flow measurements are available, on all or a subset of the network's arcs.
We divide the users into distinct groups, each group taking a different route.
The flows of all these user groups are modeled through a set of independent Poisson processes.
Our primary objective is to determine the routes taken by each user group and estimate the average rates of the corresponding Poisson processes.

\subsection{Literature review}
Our problem falls in the broader one of Network Tomography, which was formulated for the first time by Vardi in 1996 (see~\cite{VARDI1996}).
The purpose of solving this problem is to estimate the traffic demand of origin--destination (OD) pairs from the knowledge of traffic flows over some links of a network. 
Since its first appearance, the problem has been widely studied and many variations have been considered (see, e.g.,~\cite{HE2021} for a comprehensive survey on the topic).
The problem of Network Tomography is important in transportation networks, but also in data communication networks. Indeed, a large part of the relevant literature is related to this field.
For instance, Network Topology Tomography deals with the problem of reconstructing the layout of a communication network starting from a set of measurement data.
Most of Network Tomography formulations assume that a limited amount of measurements is available, and try to infer link characteristics or shared congestion in the middle of the network. For instance,~\cite{CHEN2004} addresses the problem of selecting a minimal subset of paths to be monitored with the aim of inferring the loss rates and latencies of all other paths of the network.
 
Traffic flows are usually assumed to be independent Poisson random processes; such assumption allows for a simplification of the problem formulation. In~\cite{VARDI1996}, Vardi used a second--order moment matching approach to estimate traffic demands, where a solution for the resulting linear matrix equation was obtained using an iterative minimum I--divergence procedure. 
Work~\cite{LEV-ARI2023} proposed an extension of the second--order moment matching approach of~\cite{VARDI1996} to higher order cumulants matching. The higher order cumulants introduce new useful information on the unavailable distribution of traffic flows, which can be leveraged in estimating the OD demands. By using a sufficient number of empirical cumulants, the linear mapping involved in the cumulant matching approach can achieve full column rank. In an ideal scenario where the cumulant matching equations are consistent, and the cumulants of the link traffic flows are accurately known, the rate vector can be recovered without error. It was demonstrated in~\cite{LEV-ARI2023} that this ideal situation is approachable when a sufficiently large number of realizations of traffic flows is available.
In~\cite{EPHRAIM2022}, the demands are estimated by using Moment Generating Function (MGF) matching.
This is an extension of the moment matching approach to a function that represents moments of all orders.
Finally, the Poisson assumption, although common, may constitute an oversimplification in some contexts, like internet communication networks, and can be, hence, relaxed. For instance, work~\cite{EPHRAIM2021} assumes that each OD traffic demand has a mixed Poisson distribution.

\subsection{Statement of contribution}
The contribution of this work lies in the development of a novel approach which can be applied for estimating traffic demands and roads usage in a network.
Specifically, it provides a solution to the OD Tomography problem together with that of identifying the paths with non--zero flow within a network.

In order to achieve our aim, we define a partial order on what can be regarded as the paths set, and define a monotone non--increasing function on it.
We derive some theoretical properties of the partial order and the non--increasing function.
Then, exploiting these results, we define a Dynamic Programming (DP) procedure which allows efficiently solving our estimation problem.
In this method, the paths set constitutes the state space explored by the DP. 
To the authors' knowledge, the DP approach on which the method is based is novel in the field of traffic matrix tomography and provides a new and efficient way for solving this problem.

The devised algorithm has been tested on well--known benchmark networks using synthetic data, providing a basis for comparison with existing methods which will be addressed in future works. 
Ultimately, this work introduces an advancement in traffic matrix tomography presenting an algorithm that combines a technique typical of Computer Science and Control Theory, such as DP, and an important statistical instrument, such as Cumulant Generating Functions (CGFs).

Unlike other approaches, in which only traffic demands are estimated (see, for instance,~\cite{AERDE2014,BERA2011,TEBALDI1998}) or in which the topology of the network is unknown and one wishes to reconstruct it (see, e.g.,~\cite{BERKOLAIKO2019,CACERES1999}), in this work, the topology of the network is assumed to be known a priori.
Note that such assumption is reasonable since the practical application refers to vehicular traffic networks.
Hence, the only quantities to be determined, in addition to traffic demands, are the paths that users follow.
The estimation of both these quantities can account for a more detailed analysis of traffic behavior and network usage.
It is important to remark that, if paths connecting multiple OD pairs pass through the same set of measured arcs, then we are not able to determine and separate the traffic demand contribution of each of these OD pairs individually but only their aggregate contribution.

This work also relates to~\cite{SMITH2022}, which also uses higher-order cumulants and exploits a heuristic procedure in order to reduce the number and order of the computed cumulants, which is critical in terms of numerical estimation. In our work, we also employ higher-order cumulants, however, from a theoretical point of view our method is exact in the sense that it computes deterministically all and only the cumulants needed for obtaining a solution to our problem.

\subsection{Paper organization}
The paper is organized as follows: in Section~\ref{sec:problem_formulation}, we present the estimation problem of this work in a formal algebraic setting, introduce the concept of CGFs, which is key for addressing this problem, and, under the Poisson assumption, show how their formulation can be simplified.
In Section~\ref{sec:estimation}, we discuss in further detail the addressed problems, introducing some auxiliary functions and stating some of their properties.
Furthermore, in Algorithm~\ref{alg:DP} we present the method based on DP for solving the problem of our interest.
In Section~\ref{sec:ODtomography}, we show how the problem of inferring paths and estimating demands of a traffic network falls into the class of problems presented in Section~\ref{sec:problem_formulation} and, hence, can be solved through the method introduced in Section~\ref{sec:estimation}.
In Section~\ref{sec:simulation_synthetic}, we test the devised algorithm on the widely used benchmark networks of Sioux Falls and NSFnet and show the obtained results.
Finally, in Section~\ref{sec:conclusions}, we draw the conclusions and future developments of this work.

\subsection{Notation}
We denote the set of real numbers by $\Real$. Given $X \in \Real^q$, we denote the transpose of $X$ by $X^T$. Given $A = (a_{i,j}) \in \Real^{\ell \times q}$, for $i \in \{1, \ldots, \ell\}$, we denote the $i$-th row of $A$ by $a_{i,\ast}$. For $j \in \{1, \ldots, q\}$, $X_j$ represents the $j$-th component of $X$, while $a_{\ast,j}$ is the $j$-th column of $A$. Symbols $a_{i,j}$ or $(A)_{i, j}$ denotes the $(i, j)$-th element of $A$. Let $X, Y \in \Real^q$, we denote with $X \odot Y$ the Hadamard product (or the component--wise product) of $X$ and $Y$, that is, $(\forall i \in \{1, \ldots, q\})\ (X \odot Y)_i = X_i \cdot Y_i$.

Let $X \in \Real^q$ be a vector of random variables, $E[X] = (\lambda_1, \ldots, \lambda_q)^T = \lambda$ denotes the expected value (or mean) of $X$. Such quantity is also known as the first moment of $X$. In general, we can define moments of higher order as follows, let $k \in \mathbb{N}$, a moment of order $k$ is $E[X_1^{k_1} \cdots X_q^{k_q}]$, with $k_1 + \ldots + k_q = k$.
Moreover, given a random vector $X \in \Real^q$, we define $M_X:\Real^q \rightarrow \Real$, $M_X(t) = E\left[e^{t^T X}\right]$ as the \emph{Moment Generating Function} of $X$, and consequently, $K_X: \Real^q \to \Real$, $K_X(t) = \log M_X(t)$ as the \emph{Cumulant Generating Function (CGF)} of $X$. 

A directed graph is a couple $G=(V,E)$, where $V$ is the node set and $E \subseteq V \times V$ is the edge set. We denote the number of nodes and directed edges by $n=|V|$ and $m=|E|$, respectively.
Let $s,t \in V$, a path $p$ from $s$ to $t$ is a sequence of adjacent edges without repetitions with starting node $s$ and final node $t$. 
Let $P$ be the set of all paths, we denote by $P_s^{\, t}$ the subset of $P$ consisting of those paths that connect origin $s$ to destination $t$.
We denote by $\D_P=\{(s_1,t_1),\ldots,(s_d,t_d)\} \subseteq V \times V$, with $d \in \mathbb{N}$, the pairs of starting and final nodes appearing in the paths of $P$. Note that $P = \bigcup_{(s, t) \in \D_P} P_s^{\, t}$.


\section{Problem formulation}\label{sec:problem_formulation}
Let $X=(X_1,X_2,\ldots,X_q)^T$ be a vector of $n$ real-valued mutually independent random variables with a Poisson distribution. Let $A=(a_{i,j}) \in \{0,1\}^{\ell \times q}$ be such that the columns of $A$ are all different from each other. Assume that the random vector $Y$ satisfies the linear relation
\begin{equation}\label{lin_model}
Y = AX.
\end{equation}

Since the components of random vector $X$ are independent Poisson processes, $X$ is completely defined by its mean vector $E[X]=(\lambda_1,\lambda_2,\ldots,\lambda_q)^T=\lambda$.
In our application, typically, $q$ is much larger than $\ell$.
In components, we write $Y=(Y_1,\ldots,Y_\ell)^T$.
We consider two different problems.
\begin{enumerate}
\item Assume that we know $A$ and have a full knowledge of the statistics of $Y$. We want to find $E[X]$.

\item Assume that we have a full knowledge of the statistics of $Y$, but we do not know $A$. We want to find $E[X]$ and $A$.

\end{enumerate}



\subsection{Problem 1: we know $A$ and the statistics of $Y$}

\label{sec_pois}
Let $K_Y: \Real^\ell \to \Real$ be the CGF of $Y$. Namely, for $t=(t_1,\ldots,t_\ell)^T \in \Real^\ell$, set
\[
K_Y(t) := \log M_Y(t) = \log E\left[e^{t^T Y}\right] = \log E\left[e^{t^T AX}\right].
\]
Note that
\[
t^T A X= \sum_{j=1}^{q} \left(\sum_{i=1}^\ell a_{i,j} t_i\right) X_j\,.
\]

Since the components of $X$ are independent, we can write
\[
\begin{aligned}
K_Y(t) &= \log E\left[e^{t^T A X}\right]= \log E\left[\prod_{j=1}^{q} e^{\left(\sum\limits_{i=1}^\ell a_{i,j} t_i\right) X_j}\right] = \\
&= \sum_{j=1}^{q} \log E\left[e^{\left(\sum\limits_{i=1}^\ell a_{i,j} t_i \right) X_j}\right]
=\sum_{j=1}^q C_j\left(\sum_{i=1}^\ell a_{i,j} t_i\right),
\end{aligned}
\]
where $C_j : \Real^\ell \rightarrow \Real$, with $C_j(t) = \log E[e^{X_j t}]$, is the CGF of $X_j$.
Since the components of $X$ have a Poisson distribution, the CGF of $X_j$ is
$C_j(t)= \lambda_j (e^{t}-1)$. Hence,
\[
K_Y(t) = \sum_{j=1}^q \lambda_j \left(e^{\left(\sum\limits_{i=1}^\ell a_{i,j} t_i \right)} - 1 \right),
\]
which is linear with respect to $\lambda_j$. Recall that $\lambda=(\lambda_i) \in \Real^q$. Note that
\begin{equation}\label{eq:lambda}
K_Y(t)= [f_1(t) ,\ldots,f_q(t)] \lambda,
\end{equation}
where
\[
f_j(t) = e^{\left(\sum\limits_{i=1}^\ell a_{i,j} t_i \right)} - 1.
\]
Note that~\eqref{eq:lambda} is a functional equality that needs to be satisfied for all $t \in \Real^\ell$.
In practice, one can pick a large although finite subset of $\Real^\ell$ and solve~\eqref{eq:lambda} on such set.

It remains to check if we can solve equation~\eqref{eq:lambda} with respect to $\lambda$. We can do it if and only if functions $f_1,\ldots,f_q$ are independent.
The following is a direct consequence of a Theorem due to Borel (see~\cite{BOREL1897}).
\begin{proposition}
Functions $f_1,\ldots,f_q$ are independent if and only if all columns of $A$ are different.
\end{proposition}

As a direct consequence, we have the following result.
\begin{proposition}
In model~\eqref{lin_model}, under the assumptions mentioned above, if we know the full statistics of $Y$ and matrix $A$, it is possible to compute the mean of $X$ if and only if the columns of $A$ are different.
\end{proposition}

In particular, if the columns of $A$ are different, equation~\eqref{eq:lambda} allows finding the means vector $\lambda$. Note that this result is \emph{not new}. For instance, it was already presented in~\cite{VARDI1996}, using a different argument.

\subsection{Problem 2: we only know the statistics of $Y$}

\label{sec:estimation}


We denote by $\B=\{0,1\}^\ell$ the set of logic vectors with $\ell$ components and set $\B_0 = \B \setminus \{(0, \ldots, 0)^T\}$.
Note that the cardinality of $\B$ is $|\B|=2^\ell$ and $|\B_0| = 2^\ell -1$.
We can assign a partial order (see e.g.,~\cite{DAVEY2002}) on $\B_0$ as follows.
Let $v,w \in \B_0$, with $v=(v_1,\ldots,v_\ell)^T$, $w=(w_1,\ldots,w_\ell)^T$.
Then, we set
\[
w \leq v\ \stackrel{{\text{\normalsize}}{\textrm{def}}}{\Longleftrightarrow}\ (\forall i \in \{1,\ldots,\ell\}) \ w_i \leq v_i.
\]
In other words, $w \leq v$ if each component of $w$ is lower or equal to the corresponding component of $v$.
Moreover, we write $w < v$, if $w \leq v$ and $w \neq v$.

In principle, since $\B_0$ is finite, one could apply the method presented in Section~\ref{sec_pois}, using a matrix $A$ containing all the possible $2^\ell - 1$ choices for its columns (we exclude the column of all zeros) and proceed by estimating only $E[X]$. However, this approach is possible only if $\ell$ is very small, due to the exponential growth of $|\B|$ with $\ell$. To overcome this problem, we present a method, based on DP, that allows finding the columns of $A$ without enumerating all possible elements of $\B_0$.

It is convenient to represent the columns of $A$ as elements of $\B_0$. We denote by $\A = \{a_{\ast,1}, \ldots, a_{\ast,q}\} \subseteq \B_0$ the set of columns of $A$, so that, for any $a_{\ast,j} \in \A$, with $j \in \{1, \ldots, q\}$, $X_j$ is the component of random vector $X$ associated with column $a_{\ast,j}$.
Thus, we can rewrite model~\eqref{lin_model} as
\[
Y=\sum_{j \in \{1, \ldots, q\}} a_{\ast,j} X_j.
\]

Further, we set function $\rho: \B_0 \to \Real$, such that $\rho(v)=\lambda_j$, if $(\exists j \in \{1, \ldots, q\})\ v = a_{\ast,j}$, otherwise $\rho(v) = 0$.
In other words, $\rho(v)$ is the mean of the Poisson process associated with column $v$.

We denote by $\partial_i$ the partial derivative with respect to $t_i$, and by $\partial_{i_1,\ldots,i_p}$ the partial derivative with respect to $t_{i_1},\ldots,t_{i_p}$.
 
Let $\S=(v_1,\ldots,v_r) \in \B_0^r$ be an ordered set of elements of $\B_0$. We define the logic matrix $M_{\S} \in \{0,1\}^{r \times r}$, such that
\begin{equation}\label{def:matrix_M}
(M_{\S})_{i,j}=1\ \stackrel{{\text{\normalsize}}{\textrm{def}}}{\Longleftrightarrow}\ v_i \leq v_j.
\end{equation}

\begin{proposition}\label{prop:MS}
For any non--empty $\S$, $M_{\S}$ is invertible.
\end{proposition}

\begin{proof}
Note that $(\forall i \in \{1,\ldots,r\})\ (M_{\S})_{i,i} = 1$, since for any $v \in \S$, $v \leq v$ (i.e., relation $\leq$ is reflexive).
Further, note that if $i \neq j$, at least one between $(M_{\S})_{i,j}$ and $(M_{\S})_{j,i}$ has to be zero (since relation $\leq$ is antisymmetric).
Then, matrix $M_\S$ is similar to an upper triangular matrix with all ones on the diagonal, hence, it is invertible.
\end{proof}

In particular, note that, if $\S = \A$, matrix $M_\A$ is invertible.

\begin{xmpl}\label{example:MB}
Let $\ell = 3$ and consider
\begin{equation}\label{eq:B}
\B_0 = \left\{
\stackrel{{\text{\normalsize}}\displaystyle{v_1}}{
\begin{bmatrix}
1 \\
0 \\
0
\end{bmatrix}},
\stackrel{{\text{\normalsize}}\displaystyle{v_2}}{
\begin{bmatrix}
0 \\
1 \\
0
\end{bmatrix}},
\stackrel{{\text{\normalsize}}\displaystyle{v_3}}{
\begin{bmatrix}
0 \\
0 \\
1
\end{bmatrix}},
\stackrel{{\text{\normalsize}}\displaystyle{v_4}}{
\begin{bmatrix}
1 \\
1 \\
0
\end{bmatrix}},
\stackrel{{\text{\normalsize}}\displaystyle{v_5}}{
\begin{bmatrix}
1 \\
0 \\
1
\end{bmatrix}},
\stackrel{{\text{\normalsize}}\displaystyle{v_6}}{
\begin{bmatrix}
0 \\
1 \\
1
\end{bmatrix}},
\stackrel{{\text{\normalsize}}\displaystyle{v_7}}{
\begin{bmatrix}
1 \\
1 \\
1
\end{bmatrix}}
\right\}.
\end{equation}
If we set $\S = \B_0$, we obtain the following matrix
\begin{equation}\label{example:MB}
M_{\B_0} =
\begin{bmatrix}
1 & 0 & 0 & 1 & 1 & 0 & 1 \\
0 & 1 & 0 & 1 & 0 & 1 & 1 \\
0 & 0 & 1 & 0 & 1 & 1 & 1 \\
0 & 0 & 0 & 1 & 0 & 0 & 1 \\
0 & 0 & 0 & 0 & 1 & 0 & 1 \\
0 & 0 & 0 & 0 & 0 & 1 & 1 \\
0 & 0 & 0 & 0 & 0 & 0 & 1
\end{bmatrix}.
\end{equation}
Observe that $M_{\B_0}$ is upper triangular with diagonal elements all equal to $1$ and, hence, as stated in Proposition~\ref{prop:MS}, is invertible.
It is trivial to check that $M_{\S}$ is invertible for any $\S \subseteq \B_0$.
\end{xmpl}

If $v=(v_1,\ldots,v_\ell)^T \in \B_0$, then let $i_1<\ldots<i_p \in \{1\,\ldots,\ell\}$ be such that, for $i \in \{1,\ldots,\ell\}$
$v_i=1\ \Longleftrightarrow\ i \in \{i_1,\ldots,i_p\}$.
In other words,  $i_1,\ldots,i_p$ are the indices of the components of $v$ equal to $1$.
Then, we define a function $\phi:\B_0 \to \Real$ as
\[
\phi(v) = \left.\partial_{i_1,\ldots,i_p} K_Y(t)\right|_{t = 0}.
\]

Function $\phi$ represents the \emph{joint cumulants} of variables $Y_{i_1}, \ldots, Y_{i_p}$.
The value of $\phi(v)$ corresponds to the sum of the means for all the columns $w$ of $A$ such that $v \leq w$. That is, the following property holds.
\begin{proposition}\label{prop:phi}
\[
\phi(v)=\sum_{w \in \A \mid v \leq w} \rho(w).
\]
\end{proposition}

\begin{proof}
\[
\begin{aligned}
\phi(v) & = \left.\partial_{i_1,\ldots,i_p} K_Y(t)\right|_{t = 0} =
\left.\partial_{i_1,\ldots,i_p} \left(e^{t^TA} - 1\right) \cdot\lambda \right|_{t=0} = \\
& = \left.\partial_{i_1,\ldots,i_p} \left(e^{[t_1a_{1,\ast} + \ldots + t_q a_{q,\ast}]} - 1\right) \right|_{t=0} \cdot\lambda = \\
& = \left. (a_{i_1,\ast} \odot a_{i_2,\ast} \odot \cdots \odot a_{i_p,\ast}) \odot e^{[t_1a_{1,\ast} + \ldots + t_q a_{q,\ast}]}\right|_{t=0} \cdot\lambda = \\
& = (a_{i_1,\ast} \odot a_{i_2,\ast} \odot \cdots \odot a_{i_p,\ast}) \cdot \lambda  = \sum\limits_{j = 1}^q (a_{i_1,j} \cdots a_{i_p,j})\lambda_j = \\
& = \sum\limits_{\substack{j \in \{1, \ldots, q\} \\ v \not\leq w}} (a_{i_1,j} \cdots a_{i_p,j})\lambda_j + \sum\limits_{\substack{j \in \{1, \ldots, q\} \\ v \leq w}} (a_{i_1,j} \cdots a_{i_p,j})\lambda_j = \\
& = \sum\limits_{\substack{j \in \{1, \ldots, q\} \\ v \not\leq w}} 0\cdot\lambda_j + \sum\limits_{\substack{j \in \{1, \ldots, q\} \\ v \leq w}} 1\cdot\lambda_j  = \sum_{v \leq w} \rho(w).
\end{aligned}
\vspace{-13pt}
\]
\end{proof}
 

Function $\phi$ is nonnegative and monotone non--increasing, that is, $v \leq w\ \Longrightarrow\ \phi(v) \geq \phi(w)$.

As a consequence 
\begin{equation}\label{eq:monot_non_inc}
\phi(v) = 0\ \Longrightarrow\ ((\forall w \in \B_0)\ v \leq w \Longrightarrow \phi(w)=0).
\end{equation}

Note that $\phi(v)$ can be computed by means of a sum of products of the moments of $Y$ (see, for instance, Equation (3.2.7) in~\cite{PECCATI2011}) as follows
\begin{equation}\label{eq:joint_cumulants}
\phi(v)=\!\!\!\!\sum_{\Pi \in \wp(\{1, \ldots, \ell\})}\!\!\!\! (|\Pi|-1)!(-1)^{|\Pi|-1}\prod_{\pi\in\Pi}E\left[\prod_{i\in \pi}Y_i\right].
\end{equation}

\begin{definition}
Let $v \in \B_0$. The upper set of $v$ is
\[
\uparrow\! v = \{w \in \B_0 \mid v \leq w\}.
\]
\end{definition}

\begin{xmpl}\label{example:phi}
Let us consider $\ell = 3$, $\B_0$ as in~\eqref{eq:B}, and a random vector $Y$ that satisfies model~\eqref{lin_model} with
\begin{equation}\label{eq:exampleAX}
A =
\begin{bmatrix}
1 & 0 & 0 & 1 \\
0 & 1 & 0 & 1 \\
0 & 0 & 1 & 0
\end{bmatrix}, \qquad
E[X] = \lambda= 
\begin{bmatrix}
1 \\
3 \\
2 \\
1
\end{bmatrix},
\end{equation}
and for which we know the full statistics of its components, and, in particular, the cumulants, of $Y$.
Note that, for $i \in \{1, 2, 3\}$, $\phi(v_i) = E[Y_i]$.
So, we assume to know the following quantities:
\begin{equation}\label{eq:EY}
\begin{bmatrix}
\phi(v_1) \\
\phi(v_2) \\
\phi(v_3)
\end{bmatrix}
= 
\begin{bmatrix}
E[Y_1] \\
E[Y_2] \\
E[Y_3]
\end{bmatrix}
= 
\begin{bmatrix}
2 \\
4 \\
2
\end{bmatrix},
\end{equation}
which correspond to the first order cumulants (or the means) of the components of $Y$.
We also know the second order cumulants (which correspond to the variances of the component of $Y$):
\begin{equation}\label{eq:sigmaY}
\begin{bmatrix}
\phi(v_4) \\
\phi(v_5) \\
\phi(v_6)
\end{bmatrix}
= 
\begin{bmatrix}
1 \\
0 \\
0
\end{bmatrix}.
\end{equation}
Finally, by~\eqref{eq:monot_non_inc}, we know that $\phi(v_7) = 0$, since, for instance $v_7 \in \uparrow\! v_6$.
Note that the joint cumulants of $Y$ can be computed by means of~\eqref{eq:joint_cumulants}.
Now, starting from $v_7$, down to $v_1$, we can compute $\rho$ through Proposition~\ref{prop:phi}:
$\phi(v_7) = \rho(v_7) = 0$.
Then, $\phi(v_6) = \rho(v_6) + \rho(v_7) = 0$, from which follows that $\rho(v_6) = 0$.
Analogously, $\rho(v_5) = 0$ and $\rho(v_4) = \phi(v_4) = 1$.
Consequently $\rho(v_3) = \phi(v_3) = 2$, $\rho(v_1) = \phi(v_1) - \rho(v_4) = 1$, and $\rho(v_2) = \phi(v_2) - \rho(v_4) = 3$.

\end{xmpl}

\begin{definition}
Define function $\psi:\B_0 \to \Real$ as the solution of the following recurrence equation
\[
(\forall v \in \B_0)\ \psi(v) := \phi(v) - \sum_{w \in \uparrow v \setminus v} \psi(w).
\]
\end{definition}

Note that $\psi$ satisfies the following relation:
\begin{equation}\label{eq:phi_psi}
(\forall v \in \B_0)\ \phi(v)=\sum_{w \in \uparrow v} \psi(w).
\end{equation}

\begin{proposition}
\label{prop_psi_A}
$(\forall v \in \B_0)\ \psi(v) = \rho(v)$. Moreover, if $\psi(v) \neq 0$, then $v \in \A$.
\end{proposition}

\begin{proof}
By Proposition~\ref{prop:phi} and relation~\eqref{eq:phi_psi} we have that
\begin{equation}
\label{eq:psi_phi}
(\forall v \in \B_0)\ \phi(v)=\sum_{w \in \A \mid v \leq w} \rho(w) = \sum_{w \in \uparrow v} \psi(w).
\end{equation}

We rewrite~\eqref{eq:psi_phi} as
\begin{equation}\label{eq:psi_rho}
(\forall v \in \B_0)\ \sum_{w \in \uparrow v} \left(\psi(w) - \rho(w)\right)=0.
\end{equation}

Note that, for each $v \in \B_0$,~\eqref{eq:psi_rho} can be seen as the product of the row of $M_{\B_0}$, defined as in~\eqref{def:matrix_M} over $\B_0$, associated to $v$ with column vector $[\psi(v_1) - \rho(v_1), \ldots, \psi(v_{2^\ell - 1}) - \rho(v_{2^\ell - 1})]^T$.
We can then rewrite~\eqref{eq:psi_rho} in matrix form as follows:
\[
M_{\B_0}\cdot
\left[\begin{matrix}
\psi(v_1) - \rho(v_1) \\
\vdots \\
\psi(v_{2^\ell - 1}) - \rho(v_{2^\ell - 1}) 
\end{matrix}\right]
=0.
\]
By Proposition~\ref{prop:MS}, we have that $M_{\B_0}$ is invertible.
This implies that $(\forall v \in \B_0)\ \psi(v) = \rho(v)$.
\end{proof}

So, from Proposition~\ref{prop:phi} and~\eqref{eq:psi_phi}, we have that
\begin{equation}\label{eq:solPsi}
\left[\begin{matrix}
\psi(v_1) \\
\vdots \\
\psi(v_{2^\ell-1})
\end{matrix}\right]
=
M_{\B_0}^{-1}\cdot
\left[\begin{matrix}
\phi(v_1) \\
\vdots \\
\phi(v_{2^\ell-1})
\end{matrix}\right].
\end{equation}
However, the computation of $M_{\B_0}$ and, hence, of $M_{\B_0}^{-1}$ is possible only when $\ell$ is very small.

\begin{xmpl}\label{example:psi}
Recalling Example~\ref{example:phi}, we can compute the left-hand-side of~\eqref{eq:solPsi} using $M_{\B_0}^{-1}$, with $M_{\B_0}$ as in~\eqref{example:MB}, and the values of $\phi$ over $\B_0$ as in Example~\ref{example:phi} and obtain
\[
\begin{bmatrix}
\psi(v_1) \\
\psi(v_2) \\
\psi(v_3) \\
\psi(v_4) \\
\psi(v_5) \\
\psi(v_6) \\
\psi(v_7)
\end{bmatrix}
 =
\begin{bmatrix}
1 & 0 & 0 & \!\!\!\!\!\!-1 & \!\!\!\!\!\!-1 & 0 & 1 \\
0 & 1 & 0 & \!\!\!\!\!\!-1 & 0 & \!\!\!\!\!\!-1 & 1 \\
0 & 0 & 1 & 0 & \!\!\!\!\!\!-1 & \!\!\!\!\!\!-1 & 1 \\
0 & 0 & 0 & 1 & 0 & 0 & \!\!\!\!\!\!-1 \\
0 & 0 & 0 & 0 & 1 & 0 & \!\!\!\!\!\!-1 \\
0 & 0 & 0 & 0 & 0 & 1 & \!\!\!\!\!\!-1 \\
0 & 0 & 0 & 0 & 0 & 0 & 1
\end{bmatrix} 
\cdot
\begin{bmatrix}
2 \\
4 \\
2 \\
1 \\
0 \\
0 \\
0
\end{bmatrix}
=
\begin{bmatrix}
1 \\
3 \\
2 \\
1 \\
0 \\
0 \\
0
\end{bmatrix}.
\]
As we will see in Section~\ref{sec:example_DP}, the algorithm based on DP is more efficient at estimating such values avoiding the computation of $\phi$ over all elements of $\B_0$ and, hence, computing $M_\S^{-1}$ with $\S$ in general much smaller than $\B_0$. 
\end{xmpl}

Let $v,w \in \B_0$. We say that $w$ is a successor of $v$ if $v < w$ and there are no $z \in \B_0$ such that $v<z<w$. We denote by $\rightarrow v$ the set of successors of $v$.
Note that, since $\leq$ is a partial order, we can have that $(\exists w_1, w_2 \in \B_0)\ v < w_1\ \wedge\ v < w_2\ \wedge\ w_1 \not< w_2\ \wedge\ w_2 \not< w_1$
 
Proposition~\ref{prop_psi_A} justifies the procedure presented in Algorithm~\ref{alg:DP} based on DP for finding $\A$. The algorithm traverses all elements $v \in \B_0$, for which $\phi(v)>0$. Here, $\Q$ is a FIFO queue, and $\V$ a subset of $\B_0 \times \Real$.
In what follows, set $\B_0$ is the state space of the DP procedure, and given any state $v \in \B_0$, the admissible state transitions from $v$ are all and only those to states in $\rightarrow\! v$.

\begin{algorithm}[!h]
    \begin{algorithmic}[1]
	\STATE Let $v=(0,\ldots,0)^T$. Set $\Q=\{v\}$, $\V=\varnothing$.\label{alg:step1}
	\STATE Remove from $\Q$ its first element $v$.\label{alg:step2}
	\STATE Let $\{w_1,\ldots,w_r\} =\; \rightarrow\! v$. For each $i \in \{1,\ldots,r\}$, if $\phi(w_i)\!>\!0$, add $w_i$ to $\Q$, and add pair \!$(w_i,	\phi(w_i))$ to $\V$.\label{alg:step3}
	\STATE If $\Q \neq \varnothing$, go to~\ref{alg:step2}. Otherwise the algorithm ends.\label{alg:step4}
    \end{algorithmic}
    \caption{DP for estimating $\A$ and $E[X]$}
    \label{alg:DP}
\end{algorithm}

When the algorithm terminates, let $k=|\V|$ be the cardinality of $\V = \{(w_1,\phi(w_1),$ $\ldots, (w_k,\phi(w_k))\}$. 
Set $\S=(w_1,\ldots,w_k)$.
Then, the estimated values of $x$ are given by
\[
(\psi(w_1),\ldots,\psi(w_k))^T = M_{\S}^{-1} (\phi(w_1),\ldots,\phi(w_k))^T.
\]

Finally, let $\hat \A=\{w_i \mid \psi(w_i)>0\}$.

\begin{proposition}
The previous algorithm has the following properties.
\begin{itemize}
\item[1)] $\S=\{(w,\phi(w)) \mid w \in \B_0, \phi(w)>0\}$,
\item[2)] $\hat \A=\A$ and $\psi(v)=\rho(v), v \in \A$.
\end{itemize}
\end{proposition}

\begin{proof}
\begin{itemize}
\item[1)] If $\phi(w)>0$, there exists a (strictly increasing) sequence $(0,\ldots,0)^T=v_1< \ldots < v_r=w$ such that $(\forall i \in \{2,\ldots,r\})\ v_i \in\; \rightarrow v_{i-1}$.
Moreover, since $\phi(w)>0$, being $\phi$ non--increasing, $(\forall i \in \{1,\ldots,r\})\ \phi(v_i)>0$.
By induction, note that $(v_1,\phi(v_1)) \in \S$ (step 1). Moreover, for $i \in \{1,\ldots,r-1\}$, if $(v_i,\phi(v_i)) \in \S$, $(v_{i+1},\phi_{i+1})$ is also added to $\S$, since $v_{i+1} \in \rightarrow v_i$ and $\phi(v_{i+1})>0$.
\item[2)] It follows from 1) and Proposition~\ref{prop_psi_A}.
\end{itemize}
\vspace{-15pt}
\end{proof}

\subsection{Example of Algorithm~\ref{alg:DP} execution}\label{sec:example_DP}
To clarify how Algorithm~\ref{alg:DP} works, let us consider Example~\ref{example:psi}, with $\ell = 3$, $\B_0$ defined as in~\eqref{eq:B}, and $A$ and $E[X]$ defined as in~\eqref{eq:exampleAX}.
We assume that we do not know $A$ and $E[X]$ and want to compute them from the knowledge of the cumulants of $Y$ as given in~\eqref{eq:EY} and~\eqref{eq:sigmaY}.

At Step~\ref{alg:step1}, we add $v$ to queue $\Q$ and set $\V = \varnothing$.
Then, at Step~\ref{alg:step2}, we remove $v_0$ from the queue and procede with Step~\ref{alg:step2} at computing $\rightarrow\! v_0 = \{v_1, v_2, v_3\}$ (the successors set of $v_0$).
Then, for each $i \in \{1, 2, 3\}$, if $\phi(v_i) \neq 0$, we add pair $(v_i, \phi(v_i))$ to $\V$ and element $v_i$ to queue $\Q$.
In this case, at the end of this step, we have $\Q = \{v_1, v_2, v_3\}$ and $\V = \{ (v_1, 2), (v_2, 4), (v_3, 2) \}$.
Since $\Q \neq \varnothing$, we go back to Step~\ref{alg:step2} and remove $v_1$ from the queue and compute $\rightarrow\! v_1 = \{v_4, v_5\}$.
At the end of this step $\Q = \{v_2, v_3, v_4\}$ and $\V = \{ (v_1, 2), (v_2, 4), (v_3, 2), (v_4, 1) \}$.
Note that $v_5$ has not been added to the queue since $\phi(v_5) = 0$.
Again, we remove $v_2$ from $\Q$, compute $\rightarrow\! v_2 = \{v_4, v_6\}$ and update $\Q = \{v_3, v_4\}$.
Note that at this step, we already have pair $(v_4, 1)$ in $\V$ and no new element is added to $\Q$ or $\V$ since $\phi(v_6) = 0$.
Similarly, we remove $v_3$ from $\Q$, compute $\rightarrow\! v_3 = \{v_5, v_6\}$ and note that neither $\Q$ or $\V$ need be updated.
Finally, we remove $v_4$ from $\Q$, compute $\rightarrow\! v_4 = \{v_7\}$ for which we already know that $\phi(v_7) = 0$.
Now, since $\Q = \varnothing$, Step~\ref{alg:step4} terminates the algorithm.
We can now compute $k = |\V| = 4$, set $\S = (v_1, v_2, v_3, v_4)$ and obtain $E[X]$ as follows:
$[\psi(v_1), \psi(v_2), \psi(v_3), \psi(v_4)]^T = M_\S^{-1} [\phi(v_1), \phi(v_2), \phi(v_3), \phi(v_4)]^T \Longleftrightarrow$
\[
\begin{bmatrix}
\psi(v_1) \\
\psi(v_2) \\
\psi(v_3) \\
\psi(v_4)
\end{bmatrix}
\!=\!
\begin{bmatrix}
1 & \!\! 0 & \!\! 0 & \!\! 1 \\
0 & \!\! 1 & \!\! 0 & \!\! 1 \\
0 & \!\! 0 & \!\! 1 & \!\! 0 \\
0 & \!\! 0 & \!\! 0 & \!\! 1
\end{bmatrix}^{\!\!\! -1}
\!\!\!\!\!\!\!\cdot
\begin{bmatrix}
2 \\
4 \\
2 \\
1
\end{bmatrix}
\!=\!
\begin{bmatrix}
1 & \!\! 0 & \!\! 0 & \!\!\! -1 \\
0 & \!\! 1 & \!\! 0 & \!\!\! -1 \\
0 & \!\! 0 & \!\! 1 & \!\!\! 0 \\
0 & \!\! 0 & \!\! 0 & \!\!\! 1
\end{bmatrix}
\!\cdot\!
\begin{bmatrix}
2 \\
4 \\
2 \\
1
\end{bmatrix}
\!=\!
\begin{bmatrix}
1 \\
3 \\
2 \\
1
\end{bmatrix}
\!.
\]
Finally, we observe that the vector we just computed equals $E[X]$, matrix $M_\S$ is the submatrix of~\eqref{example:MB} given by its 1st, 2nd, 3rd, and 4th columns and rows, and $(v_1, v_2, v_3, v_4) = A$.

\section{Traffic demand and path estimation from flow data}\label{sec:ODtomography}
In this section, we consider the problem of estimating traffic demands and user paths from flow measurements.
First, we show that this problem can be reduced to the one discussed in Section~\ref{sec:problem_formulation}.
Let us represent a road network by a directed graph $G=(V,E)$.
The edge set $E = \{e_1,\ldots,e_m\}$ represents the roads travelled by the users, and the node set $V=\{v_1,\ldots,v_n\}$ the intersections and endpoints.
Recall that $\D_P$ is the set of OD pairs and $P$ is the set of all paths in $G$.
Let $q = |P|$ be the cardinality of $P$. We also define the demands vector $X = (X_1, \ldots, X_q)^T$.
Vector $X$ is structured as follows: a given component $X_i$ is the flow demand associated to a given path $p \in P$.
Assuming that the traffic is not congested, we have that, for each OD pair $(s, t) \in \D_P$, all vehicles moving from $s$ to $t$ follow the shortest path in $G$ with respect to travel time.
Observe that such assumption is reasonable for road vehicles away from rush hours, or for bicycle networks.
Moreover, note that under this condition, for each $(s, t) \in \D_P$, there will be exactly one non--zero component $X_j$ associated with $(s, t)$, that is, the one associated with path $p \in P_s^{\, t}$, with $p$ being the shortest path connecting $s$ to $t$ (assuming the shortest path is unique).
For any $(s, t) \in \D_P$, the OD demand of such pair is then given by $X_j$.
Note that in case of congested traffic, for any $(s, t) \in \D_P$ there may be multiple non--zero components of $X$ associated with paths in $P_s^{\, t}$. In such case, the OD demand of such pair is given by the sum of such components (i.e., the traffic demand of a given OD pair is the sum of the traffic flows of all paths connecting that pair).
Let $T \in \mathbb{R}^m$ be a vector representing the arcs' total flows. Then, we can write
\begin{equation}\label{eq:identification}
T= MX,
\end{equation}
where $M = (m_{i,j}) \in \{0,1\}^{m \times q}$ is the edge--path incidence matrix. That is, $m_{i,j}=1$ if and only if edge $e_i$ belongs to the path associated with component $X_j$ (recalling that, for each $j \in \{1, \ldots, q\}$, there exist a path $p \in P$ associated with $X_j$).

\subsection{Problem 1: we observe all flows}\label{all_flows}
In this problem, we assume that we are able to observe traffic flows of all arcs. This means that matrix $M$ in~\eqref{eq:identification} has distinct columns. Indeed, given any path $p = (e_{i_1}, \ldots, e_{i_r}) \in P$, there exists exactly one vector of $v \in \B_0$ such that
\begin{equation}\label{edge-path}
(\forall j \in \{1, \ldots, m\})\, (v_{i_j} = 1\ \Longleftrightarrow\ e_{i_j} \in p).
\end{equation}
This means that we are in the same hypothesis of~\eqref{lin_model} and, hence, we can solve the problem of identifying $M$ and $X$ in~\eqref{eq:identification} with the DP procedure presented in Algorithm~\ref{alg:DP}.

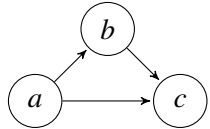
\begin{figure}[!h]
\centering
\begin{tikzpicture}[->,>=stealth',shorten >=1pt,auto,node distance=1.35cm,
style1/.style={circle, fill=none,draw=black,text=black, draw, minimum size = 20pt}]

\node[style1] (i2) {$b$};
\node[style1] (i1) [below left of=i2] {$a$};
\node[style1] (i3) [below right of=i2] {$c$};

\path (i1) edge node {} (i2)
	(i2) edge node {} (i3)
	(i1) edge node {} (i3);
\end{tikzpicture}
\caption{Graph associated with matrix $A$ of~\eqref{eq:exampleAX}.}
\label{fig:example_graph}
\end{figure}

Note that Example~\ref{example:phi}, with $\ell = 3$, $\B_0$ as in~\eqref{eq:B} and $A$ as in~\eqref{eq:exampleAX}, can be seen as a problem of form~\eqref{eq:identification} on the following graph $G = (\{a, b, c\}, \{(a, b), (a, c), (b, c)\})$, as depicted in Figure~\ref{fig:example_graph}. Indeed, $A = (v_1, v_2, v_3, v_4)$, and $v_1$, $v_2$, $v_3$, and $v_4$ correspond to paths $(a, b)$, $(b, c)$, $(a, c)$, and $((a, b), (b, c))$, respectively.

However, assuming that we can observe all traffic flows is not plausible in real--life applications. In fact, when considering real--life traffic networks only a small subset of arc flows is measured, and this motivates the following problem.

\subsection{Problem 2: we observe a subset of flows}
We now assume that we only measure a subset $\bar E = \{e_{i_1}, \ldots, e_{i_\ell}\}$ of $\ell \ll m$ arcs of $E$.
Such restriction is realistic in traffic flow measurements in which only a small set of edges flow can be retrieved.
To represent this fact, we define a projection matrix $\Pi \in \{0,1\}^{\ell \times m}$, such that $Y= \Pi T$, with $Y \in \Real^\ell$ representing the measured flows.
Moreover, by setting $A= \Pi M \in \{0,1\}^{\ell \times q}$, we obtain a linear relation similar to~\eqref{lin_model}.
However, note that after applying projection $\Pi$ to $M$, property~\eqref{edge-path} does not hold in general.
Hence, the hypothesis of~\eqref{lin_model}, which requires that the columns of $A$ are all different, is not guaranteed.
More formally, $|\A| = \bar q < q$, where $q$ is the number of columns of $A$ and $\bar q$ is the number of distinct columns of $A$.
Let us now see how we can reformulate this problem as~\eqref{lin_model}.
Let us define relation $\equiv$ in the following way: for any $p, \tilde p \in P$, let $v, \tilde v \in \B_0$ satisfy~\eqref{edge-path} for $p$ and $\tilde p$, respectively; then
\[
p \equiv \tilde p\ \stackrel{{\text{\normalsize}}{\textrm{def}}}{\Longleftrightarrow}\ \Pi(v) = \Pi(\tilde v),
\]
that is, $p$ and $\tilde p$ pass through the same set of arcs in $\bar E$.
It is easy to verify that $\equiv$ is an equivalence relation (i.e., it is reflexive, symmetric, and transitive).
This means that we can consider the quotient $\bar P = P/\equiv$ of $P$ with respect to $\equiv$.
An element $\bar p \in \bar P$ represents all paths in $P$ that pass through the same edges in $\bar E$.
Since there is a one--to--one map between $P$ and $X$, we can consider the quotient $\bar X = X/\equiv$, in which each component of $\bar X$ corresponds to one and only one element of $\bar P$.
Let $\bar q = |\bar P|$ and let $\bar A \in \{0, 1\}^{\ell \times \bar q}$ be such that its column set $\bar \A$ coincides with $\A$ (i.e., the columns of $\bar A$ are all unique and correspond to those in $\A$).
Note that $\bar X$ still satisfies the hypothesis of being a random vector of independent Poisson processes.
Indeed, each of its components is the sum of the random variables belonging to the same equivalence class and, under the Poisson assumption, this is still an independent Poisson process with mean the sum of the means of the processes of the equivalence class (see, for instance, Theorem 1.1 in Chapter 5 of~\cite{TAYLOR1998}).
Then, we can now consider the following linear relation
\[
Y = \bar A \bar X,
\]
which satisfies the hypothesis of~\eqref{lin_model}.
In this way we can exploit the method based on DP presented in Algorithm~\ref{alg:DP} for estimating traffic demands and the paths employed by the users of the network.
However, we need to better clarify what traffic demands correspond to in this context.
Note that when only $\ell < m$ arc flows are observed, one cannot retrieve the demand associated with each and every path of the network, but one can only obtain aggregated demands. More specifically, we can only estimate the demands associated with each element of quotient $\bar P$.
So we cannot obtain the single contribution of each path but only the aggregated contribution of its equivalence classes.


\section{Simulation results on synthetic data}\label{sec:simulation_synthetic}
We implemented our algorithm in Matlab. For our tests, we used a computer with an Intel(R) Core(TM) i7-4510U CPU @ 2.60 GHz processor with 16 GB of RAM. 
We tested Algorithm~\ref{alg:DP} on two graphs. The first one is not a vehicular network, but a computer networking infrastructure, known as NSFnet~\cite{BONANI2016}. The second one is a well-known traffic network, representing a part of the main roads of Sioux Falls, South Dakota, USA. 
For both networks, we assumed to know the travel times for each arc and precomputed the shortest paths for each OD pair. Then, for each test, we generated a random vector, representing the means of the independent Poisson processes $X_j$ in~(\ref{eq:identification}).

Since these tests are based on simulated data, we know exactly the means of the Poisson processes and the corresponding paths. Therefore, we can directly compare the estimated demands, obtained with our method, with the actual demands. Note that this is not the case for real--life scenarios, where flow measurements are affected by errors, and the true traffic demands and the corresponding paths are often unknown. 
In this section, we conduct tests on the problem outlined in Section~\ref{all_flows}, where we assume to measure the flow for all the arcs. For each test, we randomly generate a set of flow measurements, according to model~(\ref{eq:identification}).
We considered four different total numbers of measurements, logarithmically distributed between $1000$ and $3 \cdot 10^6$. In each test, we assumed to have only $k$ OD pairs with a flow demand higher than $0$. Moreover, we assumed that all traffic flow associated with each OD pair follows the shortest path.
We generated the flow measures as random numbers from a Poisson distribution, using the MATLAB function \textit{poissrnd}. We computed the error as the difference between the true demands and the estimated ones.

\subsection{NSFnet}
\begin{figure}[!hbt]
	\begin{center} 
\includegraphics[width=0.7\textwidth, trim=28 13 30 30, clip = true]{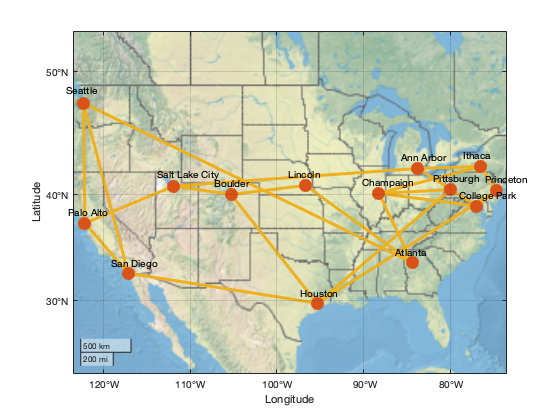}
		\caption{Graph of NSFnet.}
		\label{fig:NSFnet}
	\end{center}
\end{figure}

\begin{figure}[!htb]
	\begin{center} 
\includegraphics[width=0.7\textwidth, clip = true]{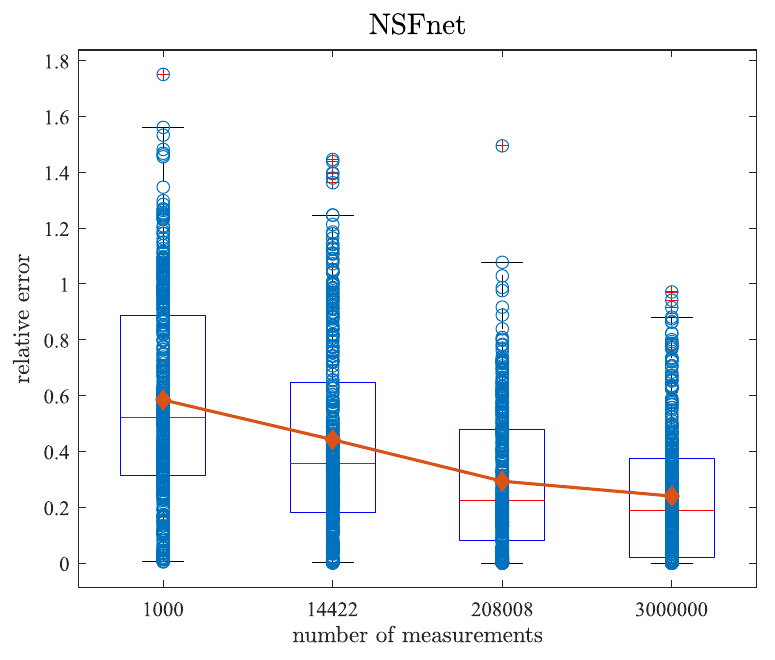}
		\caption{Relative error for 100 tests, on number of measurements. The means for each number of measurements are depicted in red.}
		\label{fig:Neterrors}
	\end{center}
\end{figure}

This network consists of $n = 14$ nodes and $m = 21$ arcs, as shown in Figure~\ref{fig:NSFnet}. Although this graph does not represent a road network, it is widely used in Network Tomography and provides a good example for small--sized tests.
Figure~\ref{fig:Neterrors} shows the error on the total flow (i.e., the sum of all demands of all OD pairs). For NSFNet, the number $k$ of OD pairs with flow demand higher than $0$ has $4$ values ($5$, $13$, $35$, $91$), logarithmically spaced. Note that the total number of OD pairs is $91$. In Figure~\ref{fig:Neterrors}, we can observe that the error decreases as the number of measurements increases. This is due to the estimation of the joint cumulants $\phi(v)$, which is more precise with a higher number of measurements.

\subsection{Sioux Falls}
\begin{figure}[!htb]
	\begin{center} 
\includegraphics[width=0.7\textwidth, trim=16 13 39 29, clip = true]{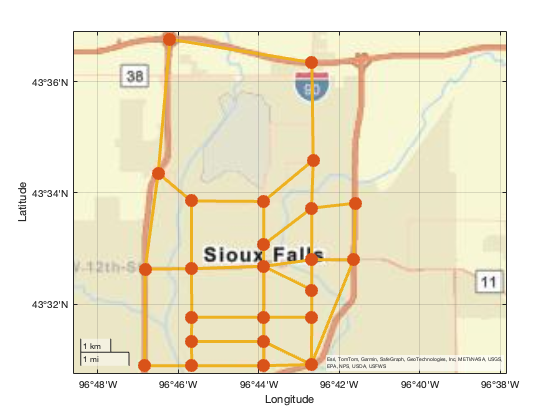}
		\caption{Graph of Sioux Falls.}
		\label{fig:SiouxFalls}
	\end{center}
\end{figure}

The Sioux Falls traffic network has been obtained from the \emph{Transportation Networks for Research} repository~\cite{TNR2024}. This network has $n = 24$ nodes and $m = 76$ arcs, as depicted in Figure~\ref{fig:SiouxFalls}. 

\begin{figure}[!htb]
	\begin{center} 
\includegraphics[width=0.7\textwidth, clip = true]{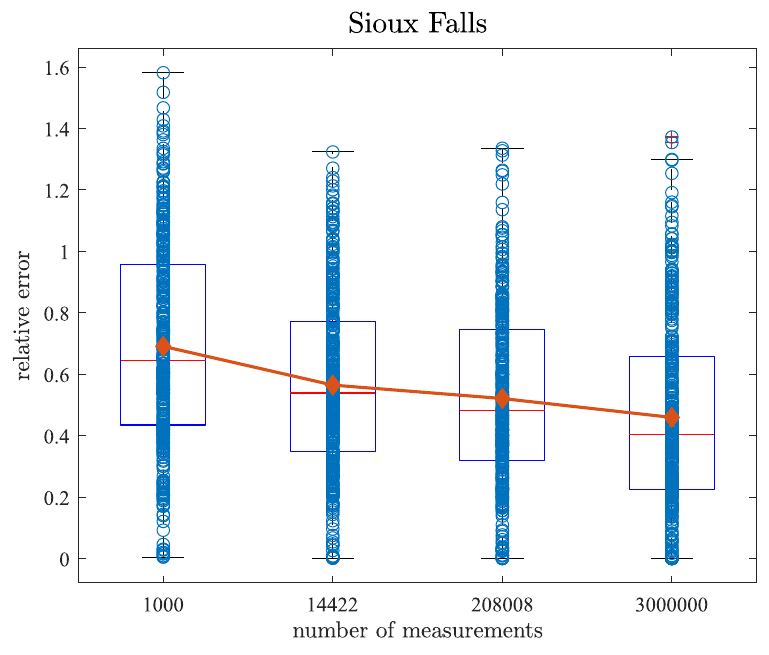}
		\caption{Relative error for 100 tests, on number of measurements. The means for each number of measurements are depicted in red.}
		\label{fig:SFerrors}
	\end{center}
\end{figure}

The number $k$ of OD pairs has $4$ values ($5$, $14$, $37$, $100$), logarithmically spaced. Figure~\ref{fig:SFerrors} reports the error divided by the total flow, for every number of measurements, with means indicated in red.    

Note that the mean error is higher compared to that on the NSFNet network. This is likely due to the fact that this graph has more arcs, and generally longer paths, than the NSFNet network.  Thus, the execution of Algorithm~\ref{alg:DP} $\phi(v)$ requires the estimation of higher order joint cumulants, which are more sensitive to outliers, and have a larger mean error.

\section{Conclusions}\label{sec:conclusions}
In this paper, we proposed a new method based on DP for both estimating the OD traffic demands and inferring the paths followed by the network users.
We described how the newly introduced algorithm exploits the use of higher-order cumulants in order to estimate the desired quantities.

We presented some theoretical results for this method and tested it on synthetic settings.
In particular, we applied the algorithm to widely used traffic networks with synthetic data.
This allowed us to test the method in a setting in which the exact solution is known.
Hence, this allowed us to compare the results obtained with our algorithm with the solutions of the synthetic tests.

Despite the theoretical discussion, in practical applications one needs to take into account some computational aspects.
In particular, the computation of cumulants of order higher than three can both be affected by numerical instability and be sensitive to noise and outliers~\cite{FILZMOSER2008}.
As a future development, we want to adapt our algorithm in order to take into account this fact by finding a good compromise between exploiting higher order cumulants while, at the same time, avoiding the drawbacks of cumulants of ``too high'' order, with the aim of making the procedure more robust.

\section*{Acknowledgments}
Project funded under the National Recovery and Resilience Plan (NRRP), Mission 4 Component 2
Investment 1.5 -- Call for tender No. 3277 of 30/12/2021 of Italian Ministry of University and Research
funded by the European Union -- NextGenerationEU.
Award Number: Project code ECS00000033, Concession Decree No. 1052 of 23/06/2022 adopted by the
Italian Ministry of University and Research, CUP D93C22000460001,
``Ecosystem for Sustainable Transition in Emilia-Romagna'' (Ecosister).

\bibliographystyle{abbrv}
\bibliography{biblio}

\end{document}